# Transition from turbulence-dominated to instability-dominated combustion regime in lean hydrogen-air flames

Hsu Chew Lee[1,2], Peng Dai[1], Minping Wan[1,2,b)], Vladimir A. Sabelnikov[3], Andrei N. Lipatnikov[4,a)]

[1]Guangdong Provincial Key Laboratory of Turbulence Research and Applications, Department of Mechanics and Aerospace Engineering, Southern University of Science and Technology, Shenzhen, Guangdong, 518055, China
[2]Guangdong-Hong Kong-Macao Joint Laboratory for Data-Driven Fluid Mechanics and Engineering Applications, Southern University of Science and Technology, Shenzhen 518055, China.
[3]ONERA – The French Aerospace Laboratory, F-91761 Palaiseau, France
[4]Department of Mechanics and Maritime Sciences, Chalmers University of Technology, Gothenburg SE-412 96
a) Author to whom correspondence should be addressed: lipatn@chalmers.se
b) wanmp@sustech.edu.cn

**Abstract**
Recent complex-chemistry direct numerical simulations of lean hydrogen-air flames propagating in forced turbulence in a box were continued by switching-off the turbulence forcing. Results show that a decrease in burning velocity $U_T(t)$, caused by the turbulence decay, is reversed when the turbulence becomes weak and a peak of $U_T(t)$ appears, with the peak magnitudes and associated Karlovitz numbers being similar in two different cases. These results (i) are attributed to activation of laminar flame instabilities, which have been suppressed by intense turbulence, and (ii) are argued to indicate that the instabilities can substantially affect $U_T$ in sufficiently weak turbulence only.

One of the greatest fundamental challenges to the combustion community consists in understanding and predicting abnormally high ratios $U_T/S_L$ of turbulent and laminar burning velocities in premixed flames characterized by a low Lewis number $Le = \kappa/D$ (here, $\kappa$ and $D$ designate molecular heat diffusivity of a mixture and molecular diffusivity of deficient reactant in this mixture, respectively). This phenomenon is well documented in (i) numerous earlier measurements reviewed elsewhere,[1,2] (ii) recent experiments with lean burning of hydrogen or $H_2$-containing fuel blends,[3-8] and (iii) recent Direct Numerical Simulation (DNS) studies.[9-12] In the literature, this phenomenon is often associated[5,6,9,10,12] with thermo-diffusive instability[13] of laminar flames characterized by a low $Le$. However, while the discussed phenomenon was documented even in highly turbulent flames[3,4,7,11] characterized by a large ratio $u'/S_L$ of rms turbulent velocity to laminar flame speed, it is not clear whether or not thermo-diffusive and hydrodynamic instabilities[13] of laminar flames play a substantial role at large $u'/S_L$. Recently, Chomiak and Lipatnikov[14] have hypothesized that such instabilities are mitigated by sufficiently intense turbulence and, hence, can substantially increase $U_T$ in weak or moderate turbulence only. More specifically, by comparing the highest strain rates created by small-scale turbulent eddies and the maximal growth rate $\omega_{max}$ of perturbations of an unstable laminar flame, Chomiak and Lipatnikov[14] have arrived at the following order-of-magnitude phenomenological criterion

$$Ka = \frac{\tau_F}{\tau_K} < Ka_{cr} = \sqrt{15}\tau_F\omega_{max} \qquad (1)$$

of importance of laminar flame instabilities in turbulent flows. Here, $Ka$ designates Karlovitz number, $\tau_F = \delta_L^T/S_L$ and $\delta_L^T = (T_b - T_u)/\max|dT/dx|$ are laminar flame time scale and thickness, respectively; $T_u$ and $T_b$ are temperatures of unburned and burned gases, respectively; $\tau_K = (\nu/\langle\varepsilon\rangle)^{1/2}$ is Kolmogorov time scale; $\nu$ is kinematic viscosity; $\langle\varepsilon\rangle = 2\langle\nu S_{ij}S_{ij}\rangle$ designates mean dissipation rate; $S_{ij} = (\partial u_i/\partial x_j + \partial u_j/\partial x_i)/2$ is the rate-of-strain tensor; and summation convention applies to repeated indexes. The present letter aims at assessing the cited hypothesis and Eq. (1) in a target-directed DNS study.

For this purpose, DNSs performed earlier by the present authors[15,16] were continued by modifying the studied cases. More specifically, the earlier three-dimensional DNSs addressed statistically one-dimensional and planar, complex-chemistry, lean (the equivalence ratio $\Phi = 0.5$, $S_L = 0.58$ m/s, $\delta_L^T = 0.41$ mm, and $\tau_F = 0.71$ ms) $H_2$-air flames propagating in forced turbulence in a box under room conditions. In the present work, results stored in two cases A and B were used as initial conditions, with the forcing being switched off and the turbulence decaying with time. These cases are characterized by $u'/S_L = 2.2$ and 4.0 (here, $u'$ is evaluated when the forcing is applied), with all other things being equal. Other details of those simulations are reported elsewhere.[15,16]

The case setup was changed to access the criterion given by Eq. (1) based on the following reasoning. If intense turbulence (i) suppresses laminar flame instabilities, as hypothesized earlier,[14] and (ii) decays with time, the instabilities could manifest themselves when the decaying turbulence becomes sufficiently weak. Therefore, a decrease in turbulent burning velocity $U_T$ with time due to the turbulence decay should be reversed, i.e., $U_T$ should increase with time, when the instabilities begin affecting the burning velocity. On the contrary, if the instabilities are not suppressed by intense upstream turbulence but always affect the turbulent flame, an increase in $U_T$ with time (due to the instability activation) should not appear.

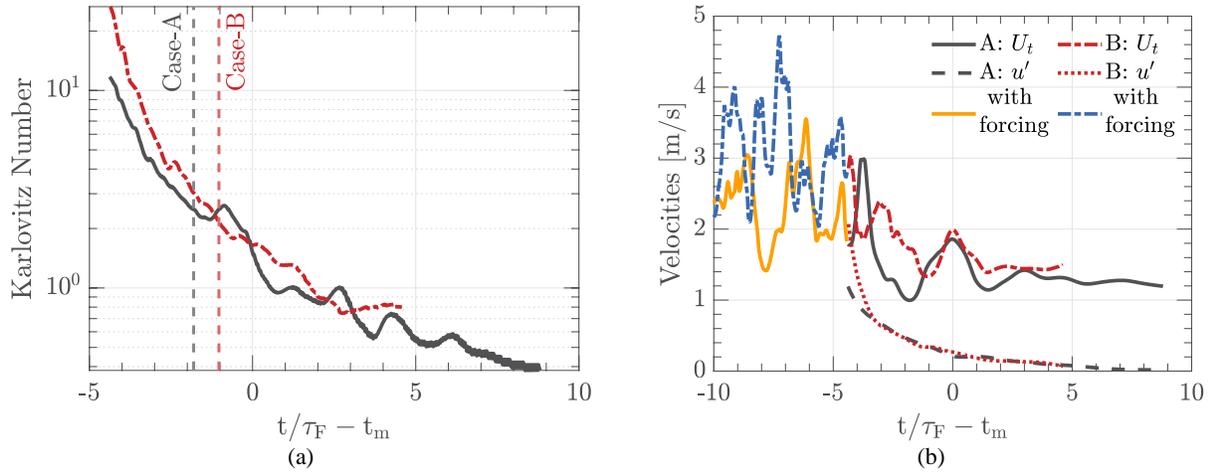

**Fig. 1.** Evolution of (a) Karlovitz number, (b) turbulent burning velocity and rms velocity. Black solid and red dotted-dashed lines show $U_T(t/\tau_F)$ obtained after switching off turbulence forcing from flames A and B, respectively. Yellow solid and blue dotted-dashed lines show $U_T(t/\tau_F)$ obtained earlier[15] in the forced cases A and B, respectively. Black dashed and red dotted lines show $u'(t/\tau_F)$. Both $Ka$ and $u'$ are averaged at the flame-brush leading edge characterized by $\langle c_F \rangle(y, t) = 0.01$. Time is normalized using flame time scale $\tau_F$ and is shifted using $t_m = 17.5$ or 25.6, which corresponds to the solitary peak of $U_T(t/\tau_F)$ in flame A or B, respectively. Vertical dashed lines show instants associated with the minima of $U_T(t/\tau_F)$ preceding these peaks.

Curves plotted in black solid and red dotted-dashed lines in Fig. 1b do show that, when the turbulence becomes weak, well-pronounced solitary peaks of $U_T(t/\tau_F)$ are observed in both cases at different normalized times $t_m$. Here, $U_T$ is evaluated by integrating normalized fuel consumption rate,[15] see Eq. (4) in the cited paper. At these instants, i.e., at $t^* \equiv t/\tau_F - t_m = 0$, the upstream rms velocity $u'$ is about 0.2 m/s in both cases, see black dashed and red dotted lines, which report its values transverse-averaged at the flame-brush leading edge, i.e., near the surface characterized by the fuel-based combustion progress variable $\langle c_F \rangle(y, t)$ close to 0.01. Karlovitz numbers transverse-averaged at the flame-brush leading edge at these instants are also almost equal in both cases, see Fig. 1a. Subsequently, at $t^* > 0$, both $Ka$ and $u'$ continue decreasing, but the burning velocity weakly oscillates around a quasi-steady value, which is close to $2.4 S_L$ in both cases.

The trends emphasized above are qualitatively consistent with the hypothesis that laminar flame instabilities are suppressed by intense turbulence but can increase burning velocity in sufficiently weak turbulence. Indeed, first, a change of the sign of the time-derivative of $U_T(t/\tau_F)$ from negative to positive, observed at $t^* \approx -2$ and $t^* \approx -1$ in cases A and B, respectively, could be attributed to transition from turbulence-dominated to instability-dominated regime of flame propagation. Second, subsequent growth of $U_T(t/\tau_F)$ to peak values observed at $t^* = 0$ could be attributed to an increase in the burning velocity due to the instability development. Third, the last stage of the $U_T(t/\tau_F)$-evolution, associated with the quasi-steady $U_T(t/\tau_F) \approx 2.4 S_L$ could be associated with the non-linear stage of the instability evolution.

The above interpretation is impeded by significant oscillations of $U_T(t/\tau_F)$, computed in the original forced-turbulence cases A and B, see curves plotted in yellow solid and blue dotted-dashed lines in Fig. 1b. These curves report only a part of DNS data published earlier,[15,16] as results obtained at $t/\tau_F < 7.5$ (case A) or $t/\tau_F < 15.6$ (case B) are omitted in Fig. 1b to place the focus of consideration on the new DNS data simulated in the case of decaying turbulence. Similar oscillations of $U_T(t)$ were observed in earlier DNS studies of various turbulent premixed flames[17-19] and the reader interested in physical mechanisms that cause these oscillations is referred to the cited papers. To support the hypothesis discussed here, we may note that the solitary peaks observed at $t^* = 0$ in Fig. 1b (i) are characterized by a significantly smaller height-to-width ratio when compared to the preceding peaks and (ii) are associated with the end of the decreasing trend for $U_T(t/\tau_F)$.

Nevertheless, these features of the highlighted peaks are not sufficient to claim that the DNS data plotted in Fig. 1 prove the hypothesis. Therefore, since the present authors are not aware on a method of analysis of DNS data, which could yield decisive evidence as to whether turbulence or laminar flame instability is of more importance, let us consider DNS data that support the studied hypothesis and Eq. (1) indirectly.

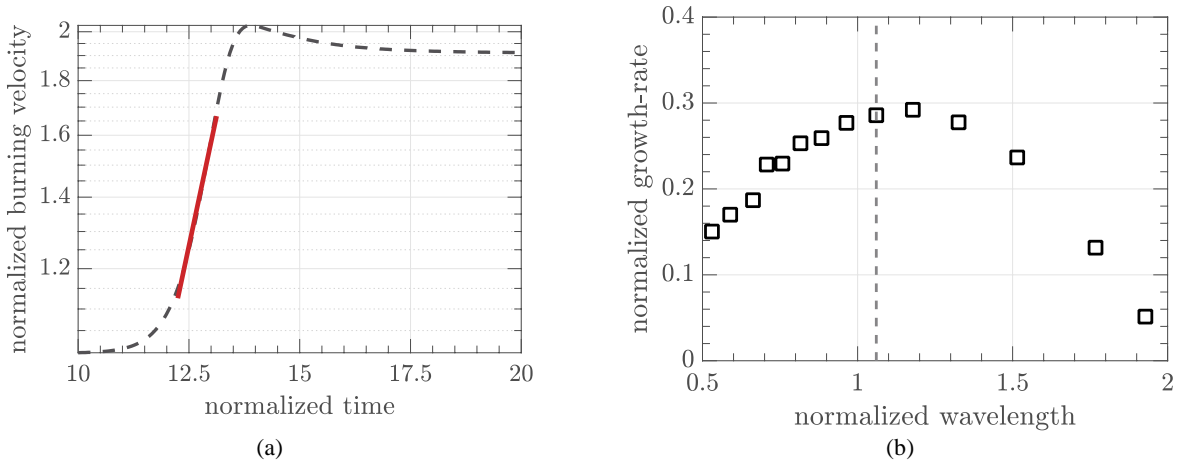

**Fig. 2.** (a) Growth of normalized laminar burning velocity $U_L/S_L$ in a three-dimensional case ($\Lambda = 2.4$ mm). Red straight line shows a linear fit to $ln(U_L/S_L)$ vs. $t/\tau_F$. (b) Normalized instability growth rate $\tau_F \omega$ vs. normalized wavenumber $\delta_L^T k$ in two-dimensional case. Vertical dashed line shows wavenumber adopted in the three-dimensional case.

To obtain such data, two and three-dimensional simulations of unstable laminar flames were run using the same software, the same chemical mechanism, and the same molecular transport model. The three-dimensional simulations were performed in the same computational domain, whereas its width $\Lambda$ was varied in the two-dimensional simulations. In all cases, a small periodic perturbation of the axial velocity was set at the inlet boundary $y = 0$, with the perturbation wavelength being equal to $\Lambda$. In unstable cases, the computed burning velocity grew exponentially with time during the linear stage of the instability development and the instability growth rate $\omega(k)$ was evaluated using linear fit to the computed dependencies of $ln(U_L/S_L)$ on time, e.g., see red straight line in Fig. 2a, which yields $\tau_F \omega = 0.48$. Here, $k = 2\pi/\Lambda$ is the instability wavelength and $U_L$ designates burning velocity obtained from an unstable laminar flame. It is worth stressing that those simulations allowed for both thermo-diffusive and hydrodynamic instabilities.[13]

Dispersion curves $\tau_F \omega(\delta_L^T k)$ obtained from two-dimensional flames and plotted in Fig. 2b are similar to dispersion curves computed by other research groups in two-dimensional complex-chemistry simulations of lean hydrogen-air laminar flames.[20,21] The following quantitative characteristics of the studied unstable laminar flames are relevant to the goal of the present work.

First, at large $t/\tau_F$, the ratio $U_L/S_L$ is equal to 1.9 in Fig. 2a, whereas the quasi-steady value of $U_T$ is about $2.4 S_L$ in Fig. 1b. These two numbers are sufficiently close to one another, thus, indirectly supporting an association of the quasi-steady $U_T$ with the non-linear stage of the instability development. Difference in the two numbers is not surprising, because turbulence does not decay completely in case A or B and could contribute to increasing flame surface area and $U_T$. Moreover, Fig. 2a shows evolution of laminar flame instabilities initiated by a weak periodic perturbation with a single wavelength, whereas the quasi-steady stage in Fig. 1b results from evolution of multiscale perturbations created by turbulence. However, even if the quasi-steady stage is controlled by laminar flame instabilities, this does not mean that the instabilities were suppressed by more intense turbulence. Nevertheless, comparison of $U_L/S_L = 1.9$ and $U_T/S_L = 3.3$ obtained from flame A at the onset of turbulence decay ($t^* = -4.5$) implies that an increase in $U_T$ due to turbulence is larger than an increase in $U_T$ due to the instabilities even in case A. Finally, it is worth stressing that the discussed numbers (1.9 and 2.4) are specific to the considered computational domain whose width $\Lambda = 2.4$ mm. In wider computational domains, dynamics of unstable laminar flames is substantially richer and larger ratios $U_L/S_L$ can be reached.[22-24] Moreover, when the size of unstable laminar flames was sufficiently large, continuous self-acceleration of the flames due to its hydrodynamic instability was documented.[25] These phenomena are beyond the scope of the present study whose focus is solely placed on eventual suppression of laminar flame instabilities by intense small-scale turbulence.[14]

Second, by redrawing the dependencies of $U_T(t/\tau_F)$ plotted in Fig. 1b in the logarithmic scale and adapting the method illustrated in Fig. 2a, the maximal growth rates were calculated at $-2 < t^* < 0$ and $-1 < t^* < 0$ in cases A and B, respectively. These maximal growth rates normalized using $\tau_F$ are equal to 0.67 and 0.58, respectively, and are sufficiently close to $\tau_F \omega = 0.48$ obtained from the three-dimensional unstable laminar flame in the same computational domain, see Fig. 2a. Again, differences in the turbulent and laminar growth rates are not surprising for the reasons discussed above.

Third, comparison of Figs. 2a and 2b implies that the maximal instability growth rate is close to $\tau_F \omega_{max} = 0.5$ in the three-dimensional case in the computational domain adopted in the DNSs of decaying turbulence. Substitution of this number into Eq. (1) yields $Ka_{cr} = 1.9$. To test this criterion, Karlovitz numbers evaluated at instants marked with straight dashed lines in Fig. 1a could be adopted. Indeed, at these instants, $U_T(t/\tau_F)$ reaches minimal values, which could be attributed to a balance between a decrease in the burning velocity due to turbulence decay and an increase in $U_T$ due to activation of laminar flame instabilities, which have earlier been suppressed by more intense turbulence. Figure 1a shows that, at these instants, $Ka \approx 2$ and 2.3 in cases A and B, respectively. These DNS data are close to $Ka_{cr} = 1.9$ yielded by Eq. (1), thus, supporting this criterion. This support appears to be sufficiently strong because Eq. (1) was obtained before the DNSs considered in the present work were started in the case of decaying turbulence.

Finally, to further support the discussed hypothesis from the qualitative perspective, Fig. 3 shows images of the instantaneous flame surfaces $c_F(\mathbf{x}, t) = c^*$ sampled from flame B at three instants: (a) turbulence forcing is switched off, i.e., $t^* \approx -4.2$, (b) the minimum of $U_T(t/\tau_F)$ preceding the peak $U_T(t^* = 0)$, i.e., $t^* \approx -1$, and (c) $t^* = 0$. Here, $c^*$ refers to $c_F(\mathbf{x}, t)$ corresponding to the peak Fuel Consumption Rate (FCR) in the turbulent flame brush. At $t^* \approx -4.2$, the flame surface is complicated, see Fig. 3a, with spatial scales of the surface wrinkles being significantly smaller than the wavelength $\Lambda$ associated with the peak instability growth rate, see Fig. 2b. At $t^* \approx -1$, see Fig. 3b, the flame-surface is much more regular and large-scale perturbations whose lengths are comparable with $\Lambda$ are well pronounced. At $t^* = 0$ associated with the peak $U_T(t/\tau_F)$, the flame-surface looks basically similar. Thus, comparison of Figs.

3a and 3b or 3c indicates transition from turbulence-dominated to instability-dominated regime of flame propagation, with no manifestation of the instabilities being observed in sufficiently intense turbulence, see Fig. 3a.

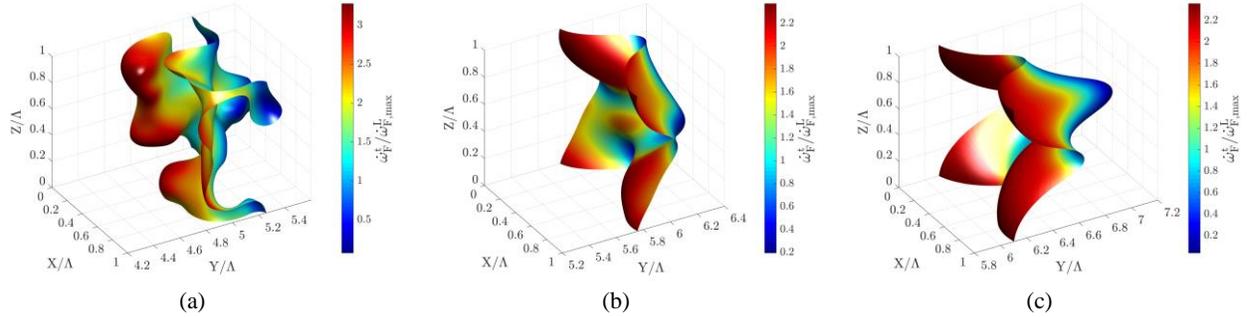

**Fig. 3.** Images of instantaneous B-flame surfaces $c_F(\mathbf{x},t) = c^*$ at (a) $t^* \approx -4.2$, (b) $t^* \approx -1$, and (c) $t^* = 0$. Color scales show the local FCR normalized using the maximum FCR in the unperturbed laminar flame.

In summary, the performed DNS study not only provides indirect evidence of (i) the hypothesis of suppression of laminar flame instabilities (both thermo-diffusive and hydrodynamic ones) by sufficiently intense turbulence and (ii) transition from turbulence-dominated to instability-dominated regime of flame propagation, but also validates the criterion given by Eq. (1) and based on this hypothesis.[14] While more research into the issue is definitely required, the reported evidence and the quantitative validation of Eq. (1) should not be ignored and call for further target-directed studies.

As far as large ratios of $U_T/S_L$, well-documented in highly turbulent lean $H_2$-air flames, are concerned, this phenomenon is associated with a significant increase in local burning rate in highly curved reaction zones localized to the leading edges of turbulent flame brushes.[1,2] The present DNS data do show that such an increase is more pronounced in more intense turbulence, cf. color scales in Figs. 3a and 3b or 3c. The reader interested in the latest progress in this research direction is referred to recent papers.[26-31]

**Declaration of Competing Interest**
The authors declare that they have no known competing financial interests or personal relationships that could have appeared to influence the work reported in this paper.

**Acknowledgements**
ANL gratefully acknowledges the financial support provided by Chalmers Area of Advance Transport. VAS gratefully acknowledges the financial support provided by ONERA. Other authors have been supported in part by NSFC (Grant Nos. 91752201, 51976088, and 92041001), the Shenzhen Science and Technology Program (Grant No. KQTD20180411143441009 and JCYJ20210324104802005), Department of Science and Technology of Guangdong Province (Grant Nos. 2019B21203001 and 2020B1212030001), National Science and Technology Major Project (Grant Nos. J2019-II-0006-0026 and J2019-II-0013-0033), and the Center for Computational Science and Engineering of SUSTech.